\documentclass[aps,prd,showpacs]{revtex4}

\usepackage{amsfonts}
\usepackage{verbatim}
\usepackage{amsmath}
\usepackage[USenglish]{babel}
\usepackage[dvips]{epsfig}
\usepackage{graphicx}
\usepackage{float}
\usepackage{epsfig,subfig}
\usepackage{epstopdf}
\usepackage{hyperref}
\usepackage{cleveref}
\usepackage{color}
\DeclareMathAlphabet\mathbfcal{OMS}{cmsy}{b}{n}
\usepackage{lipsum}

\begin{document}

\title{Local effects of the quantum vacuum in Lorentz-violating electrodynamics}

\author{A. Mart\'{i}n-Ruiz}
\email{alberto.martin@nucleares.unam.mx}
\affiliation{Instituto de Ciencias Nucleares, Universidad Nacional Aut{\'o}noma de M{\'e}xico, 04510 M{\'e}xico, Distrito Federal, M{\'e}xico.}

\author{C. A. Escobar}
\email{carlos.escobar@correo.nucleares.unam.mx}
\affiliation{Departamento de F\'isica, Universidade do Algarve, 8005-139 Faro, Portugal.}

\begin{abstract}
The Casimir effect is one of the most remarkable consequences of the non-zero vacuum energy predicted by quantum field theory. In this paper we use a local approach to study the Lorentz violation effects of the minimal standard model extension on the Casimir force between two parallel conducting plates in the vacuum. Using a perturbative method similar to that used for obtaining the Born series for the scattering amplitudes in quantum mechanics, we compute, at leading order in the Lorentz-violating coefficients, the relevant Green's function which satisfies given boundary conditions. The standard point-splitting technique allow us to express the vacuum expectation value of the stress-energy tensor in terms of the Green's function. We discuss its structure in the region between the plates. We compute the renormalized vacuum stress, which is obtained as the difference between the vacuum stress in the presence of the plates and that of the vacuum. The Casimir force is evaluated in an analytical fashion by two methods: by differentiating the renormalized global energy density and by computing the normal-normal component of the renormalized vacuum stress. We compute the local Casimir energy, which is found to diverge as approaching the plates, and we demonstrate that it does not contribute to the observable force.
\end{abstract}

\maketitle

\section{Introduction}

Symmetry principles play a fundamental role in theoretical physics. As an outstanding example, Lorentz invariance is one of the cornerstones of general relativity (GR) and the standard model (SM) of particle physics. Although both theories have been successful in explaining and predicting the observed physical phenomena with a high degree of accuracy, they come with their own set of deficiencies: ultraviolet divergences in quantum field theories and singularities in general relativity. It is generally believed that an unified quantum theory of gravity will solve these notable problems; and thus its search has become one of the most important challenges of theoretical physicists.

Interest in Lorentz violation \cite{Noconmu,Bworld,emergentG,modelos1,modelos2,modelos3} has grown rapidly in the last decades since many candidate theories of quantum gravity \cite{quantum1,quantum2}, such as string theory \cite{String2} and loop quantum gravity \cite{QG1}, possess scenarios involving deviations from Lorentz symmetry. Nowadays, investigations concerning Lorentz violation are mostly conducted under the framework of the standard model extension (SME), initiated by Kosteleck\'{y} and Colladay \cite{SME,SME2}. The SME is an effective field theory that contains the standard model, general relativity, and all possible operators that break Lorentz invariance. The Lorentz-violating (LV) coefficients arise as vacuum expectation values of some basic fields belonging to a more fundamental theory, such a string theory \cite{Strings}. Some important features of the minimal SME comprise invariance under observer Lorentz transformations, energy-momentum conservation, gauge invariance, power-counting renormalizability \cite{Ren1,Ren2,Ren3}, causality, stability and hermiticity \cite{Causality}.

A Lorentz-violating vacuum acts in many respects like a nontrivial optical medium. Therefore, one expects the photon sector of the SME to possess features similar to those of ordinary electrodynamics in macroscopic media. The electrodynamics limit of the SME has been widely studied in the literature \cite{Light1}. Indeed, changes in the propagation of light, such as polarization, birefringence \cite{Light1,Light2,Light3} and \v{C}erenkov effect \cite{Cherenkov1,Cherenkov2,Cherenkov3}, have been predicted and used to place tight bounds on Lorentz violation. The main goal of this work is to provide additional contributions regarding the local effects of the quantum vacuum in a particular sector of the electrodynamics limit of the SME, namely, the CPT-odd Maxwell-Chern-Simons term \cite{CFJ}. Concretely, we study the Casimir effect (CE) between two parallel conducting plates using a local approach based on the calculation of the vacuum expectation value of the stress-energy tensor via Green's functions satisfying the suitable boundary conditions.

In its simple manifestation, the CE is a quantum force of attraction between two parallel uncharged conducting plates \cite{Casimir}. More generally, it refers to the stress on bounding surfaces when a quantum field is confined to a finite volume of space. The boundaries can be material media, interfaces between two phases of the vacuum, or topologies of space. In any case, the modes of the quantum fields are restricted, giving rise to a macroscopically measurable force \cite{Bressi}. The CE has been studied in different scenarios, including the standard model, the gravitational sector \cite{Quach}, the AdS-CFT correspondence \cite{AdS}, condensed matter systems \cite{CE-TI2,CE-TI} and chiral metamaterials \cite{CE-MM}, to name a few. 

The CE has also been considered within the SME framework \cite{Martin-Escobar, Mariana,Kharlanov}. The authors in Ref. \cite{Kharlanov} used the zeta function regularization technique to compute the Casimir force between two parallel conducting plates within the (3+1)D Maxwell-Chern-Simons theory. As a consistency check, they also evaluated the renormalized vacuum energy by a series summation (via the residue theorem) of one-particle energy eigenstates. The first attempt to tackle this problem was due to M. Frank and I. Turan \cite{Mariana}; however, as pointed out by O. G. Kharlanov and V. Ch. Zhukovsky \cite{Kharlanov}, they used misinterpreted equations which led to an oversimplified treatment of the problem. More precisely, they considered that the photon dispersion relation corresponds to that for a massive photon; however, unlike the (2+1)D case, in (3+1)D the effect of the Maxwell-Chern-Simons term is a more complicated dispersion relation for the photon. Due to this wrong equation, Frank and Turan constructed also incorrectly the relevant Green's function (GF). One of the specific aims of this work is the construction of the correct Green's function within the (3+1)D Maxwell-Chern-Simons theory and the calculation of the Casimir energy density and stress between two parallel conducting plates. Furthermore, the method can be further generalized to diverse geometries. The basics for the construction of the GF is that the Lorentz-violating field equations can be treated perturbatively due to the smallness of the LV coefficients \cite{cotas}. Our Green's functions also provide information about the divergence of the local energy density near the plates.

The outline of this paper is as follows. \Cref{model} reviews some basics of the particular sector of the minimal SME to be considered in this work, namely, the (3+1)D Maxwell-Chern-Simons model. Using a perturbative method similar to that used for obtaining the Born series for the scattering amplitudes in quantum mechanics, in \cref{GF-method} we compute the leading-order Green's function which satisfies given Dirichlet, Neumann or Robin boundary conditions, provided the smallness of the LV coefficients. In \cref{VS-section} we use the standard point-splitting technique to express the vacuum expectation value of the stress-energy tensor in terms of the Green's function. The concrete calculation of the renormalized vacuum stress (and the Casimir force) between two parallel conducting plates is performed in \cref{CasEff-Section}. We also discuss the local energy density, which is found to diverge as approaching the plates. We demonstrate that the divergent term does not contribute to the observable force. The conclusions are contained in \cref{conclusions}. Details of technical computations are left to the appendix. Here, Lorentz-Heaviside units are assumed ($\hbar =c=1$), the metric signature will be taken as $\left( +,-,-,-\right) $ and the convention $\epsilon ^{0123}=+1$ is adopted.

\section{Lorentz-violating electrodynamics} \label{model}

The renormalizable gauge-invariant photon sector of the SME consists of the usual Maxwell Lagrangian plus the additional terms $\frac{1}{2} (k _{AF}) ^{\kappa} \epsilon _{\kappa \lambda \mu \nu} A ^{\lambda} F ^{\mu \nu}$ and $- \frac{1}{4}(k _{F}) _{\alpha \beta \mu \nu} F ^{\alpha \beta} F ^{\mu \nu}$. The LV tensor coefficients $(k _{AF}) ^{\kappa}$ and $(k _{F}) _{\alpha \beta \mu \nu}$ are CPT-odd and CPT-even, respectively. Many components of these coefficients are strongly constrained by astrophysical spectropolarimetry \cite{CFJ}. Despite this, further investigations remain to be of great interest both for a better understanding of massless Lorentz-violating fields and for the potential complementary tighter bounds.

In this paper we are concerned with the CPT-odd sector. The relevant Lagrangian is
\begin{align}
\mathcal{L} =  - \frac{1}{4} F _{\mu \nu} F ^{\mu \nu} + ( k _{AF} ) _{\mu} A _{\nu} \tilde{F} ^{\mu \nu} - j _{\mu} A ^{\mu} . \label{Lagrangian}
\end{align}
Here, $j ^{\mu} = ( \rho, \mathbf{J} )$ is the 4-current source that couples to the electromagnetic 4-potential $A ^{\mu}$, $F ^{\mu \nu} = \partial ^{\mu} A ^{\nu} - \partial ^{\nu} A ^{\mu}$ is the electromagnetic field strength and $\tilde{F} ^{\mu \nu} = \frac{1}{2} \epsilon ^{\mu \nu \alpha \beta} F _{\alpha \beta}$ its dual. Since we take $(k _{F}) _{\mu \nu \alpha \beta} = 0$, we can omit the subscript $AF$ of the Lorentz- and CPT-violating $( k _{AF} ) ^{\mu}$ coefficients and set $( k _{AF} ) ^{\mu} \equiv k ^{\mu} = (k ^{0} , \textbf{k})$. A nondynamical fixed $k ^{\mu}$ determines a special direction in spacetime. For example, certain features of plane wave propagating along $\textbf{k}$ might differ from those of waves perpendicular to $\textbf{k}$. Thus, particle Lorentz transformations are violated.

Varying the action $\mathcal{S} = \int \mathcal{L}\, d ^{4} x$ with respect to $A ^{\mu}$ yields the equations of motion for the 4-potential $A ^{\mu} = (A ^{0} , \textbf{A})$:
\begin{align}
\left( \Box \eta ^{\mu} _{\phantom{\mu} \nu} - \partial ^{\mu} \partial _{\nu} - 2 k _{\beta} \epsilon ^{\mu \beta \alpha} _{\phantom{\mu \beta \alpha} \nu} \partial _{\alpha} \right) A ^{\nu} = j ^{\mu} , \label{Field-Eqs}
\end{align}
which extend the usual covariant Maxwell equations to incorporate Lorentz violation. Of course, the homogeneous Maxwell equations that express the field-potential relationship
\begin{align}
\partial _{\mu} \tilde{F} ^{\mu \nu} = 0 \label{Bianchi}
\end{align}
are not modified due to the $U(1)$ gauge invariance of the action. As in conventional electrodynamics, current conservation $\partial _{\mu} j ^{\mu} = 0$ can be verified directly by taking the divergence at both sides of Eq. (\ref{Field-Eqs}). 

In noncovariant notation, the inhomogeneous equations of motion (\ref{Field-Eqs}) reads
\begin{align}
\label{MaxEqs}
\begin{split}
\nabla \cdot \textbf{E} - 2 \textbf{k} \cdot \textbf{B} = \rho , \\[7pt] \nabla \times \textbf{B} - \dot{\textbf{E}} - 2 k ^{0} \textbf{B} + 2 \textbf{k} \times \textbf{E} = \textbf{J} ,
\end{split}
\end{align}
while the homogeneous equations (\ref{Bianchi}) are $\nabla \cdot \textbf{B} = 0$ and $\nabla \times \textbf{E} + \dot{\textbf{B}} = 0$. Gauge invariance of physics is evident from Eqs. (\ref{MaxEqs}), and any of the usual conditions on $A ^{\mu}$, like Lorentz or Coulomb gauge, can be imposed. Interestingly, the microscopic equations (\ref{MaxEqs}) can be cast in the form of Maxwell equations for macroscopic media:
\begin{align}
\nabla \cdot \textbf{D} = \rho \qquad , \qquad \nabla \times \textbf{H} - \dot{\textbf{D}} = \textbf{J} , \label{MaxEqsMatter}
\end{align}
with the modified constitutive relations
\begin{align}
\textbf{D} = \textbf{E} + \theta (\textbf{x} , t) \; \textbf{B} \quad , \quad \textbf{H} = \textbf{B} - \theta (\textbf{x} , t) \; \textbf{E} , \label{ConstRelations}
\end{align}
where $\theta (\textbf{x} , t) = 2 k _{\mu} x ^{\mu} = 2 k ^{0} t - 2 \textbf{k} \cdot \textbf{x}$ is a spacetime dependent axion field. These equations reveal a remarkable feature of this theory, the magnetoelectric effect, which also occurs in the CPT-even photon sector of the SME \cite{Martin-Escobar}. Furthermore, Eqs. (\ref{MaxEqsMatter}) together with the constitutive relations of Eq. (\ref{ConstRelations}) resemble to those describing the electromagnetic response of condensed matter systems, in which case, the spatial and temporal dependence of the axion field defines a specific realization of a topologically-nontrivial phase of matter. For example, $\theta = \pi$ in the case of 3D topological insulators \cite{TI}, and $\theta (\textbf{x} , t) = 2 b ^{0} t - 2 \textbf{b} \cdot \textbf{x}$ for Weyl semimetals \cite{WSM}, where the parameter $2 b ^{0}$ is interpreted as the separation between the nodes in energy, and $2 \textbf{b}$ denotes the separation between the Weyl nodes in momentum space. However, note that for condensed matter systems, $\theta$ is defined inside the material, while in the SME framework $k _{\mu}$ permeates the whole spacetime.

The stress-energy tensor for this theory is given by
\begin{align}
\Theta ^{\mu \nu} = - F ^{\mu \alpha} F ^{\nu} _{\phantom{\nu} \alpha} + \frac{1}{4} \eta ^{\mu \nu} F ^{\alpha \beta} F _{\alpha \beta} - k ^{\nu} \tilde{F} ^{\mu \alpha} A _{\alpha} . \label{EM-tensor}
\end{align}
Here $\eta ^{\mu \nu} = \mbox{diag} (1,-1,-1,-1)$ denotes the usual Minkowski flat space-time metric. Unlike the conventional case, $\Theta ^{\mu \nu}$ cannot be symmetrized because its antisymmetric part is not longer a total derivative. By virtue of the equations of motion (\ref{Field-Eqs}) and (\ref{Bianchi}), the energy-momentum tensor obeys
\begin{align}
\partial _{\mu} \Theta ^{\mu \nu} = j _{\mu} F ^{\mu \nu} , \label{Div-EM-tensor}
\end{align}
which implies that it is conserved in the absence of sources. Although the energy-momentum tensor is gauge dependent, it only changes by a total derivative under the gauge transformation $A ^{\mu} \rightarrow A ^{\mu} + \partial ^{\mu} \Lambda$, i.e.
\begin{align}
\tilde{F} ^{\mu \alpha} A _{\alpha} \rightarrow \tilde{F} ^{\mu \alpha} A _{\alpha} - \partial _{\alpha} (\tilde{F} ^{\mu \alpha} \Lambda) .
\end{align}
Consequently, the integrals over all space are gauge invariant. Note that the energy
\begin{align}
\mathcal{E} = \int \Theta ^{00} d ^{3} \textbf{x} = \int \frac{1}{2} \left( \textbf{E} ^{2} + \textbf{B} ^{2} - k ^{0} \textbf{B} \cdot \textbf{A} \right) d ^{3} \textbf{x}
\end{align}
 is not positive definite due to the term $k ^{0} \textbf{B} \cdot \textbf{A}$, which may be negative. The appearance of this term in the energy density can introduce instability in the theory, and it can be resolved by requiring that only spacelike components of $k ^{\mu}$ are nonzero. However, this condition depends on the observer frame, so even an infinitesimal boost to another observer frame would reintroduce instability. Despite arising from a hitherto unobserved spontaneous breaking of the electromagnetic $U(1)$ gauge symmetry, the photon mass can be introduced in this theory to eliminate the linear instability. Although this idea might be physically acceptable, in this work we restrict ourselves to the minimal modification of the usual standard model.

\section{Green's function method} \label{GF-method}

Knowledge of Green's function (GF) allows one to compute the electromagnetic fields for an arbitrary distribution of sources, as well as to solve problems with given Dirichlet, Neumann or Robin boundary conditions on arbitrary surfaces. To derive the GF for the previously discussed LV electrodynamics one can employ standard Fourier methods. As in conventional electrodynamics, the modified Maxwell operator appearing in parentheses in Eq. (\ref{Field-Eqs}) is singular, as one can verify in Fourier space. To circumvent the non invertibility of the corresponding Minkowski matrix one can further work in Lorentz gauge. The free-space GF (satisfying the standard boundary conditions at infinity) in momentum \cite{Cherenkov3} and coordinate \cite{brasilenos} representations can be obtained in a simple fashion. In this paper we are concerned with the effects of this Lorentz-violating electrodynamics on the Casimir force between two parallel conducting plates in the vacuum. To this end we employ a local approach consisting in the evaluation of the vacuum expectation value of the stress-energy tensor of the system, which can be expressed in terms of the appropriate Green's function. The presence of boundaries (e.g. the plates) makes the GF derived in Refs. \cite{Cherenkov3,brasilenos} not suitable for our purposes. Thus the aim of this section is the construction of the Green's functions which incorporates the presence of boundaries.

In the Lorentz gauge $\partial _{\mu} A ^{\mu} = 0$, the field equations (\ref{Field-Eqs}) take the form
\begin{align}
\left( \Box \eta ^{\mu} _{\phantom{\mu} \nu} - 2 k _{\beta} \epsilon ^{\mu \beta \alpha} _{\phantom{\mu \beta \alpha} \nu} \partial _{\alpha} \right) A ^{\nu} = j ^{\mu} , \label{Field-Eqs2}
\end{align}
where $\Box = \partial _{\mu} \partial ^{\mu} = \partial ^{2} _{t} - \nabla ^{2}$ is the D'Alambert operator. To obtain the general solution of Eq. (\ref{Field-Eqs2}) for arbitrary external sources, we introduce the GF matrix $G ^{\mu} _{\phantom{\mu}\nu} (x,x^{\prime})$ solving Eq. (\ref{Field-Eqs2}) for a pointlike source,
\begin{align}
\left( \Box \eta ^{\mu} _{\phantom{\mu} \nu} - 2 k _{\beta} \epsilon ^{\mu \beta \alpha} _{\phantom{\mu \beta \alpha} \nu} \partial _{\alpha} \right)  G ^{\nu} _{\phantom{\nu} \gamma} (x , x ^{\prime}) = \eta ^{\mu} _{\phantom{\mu} \gamma} \delta ^{4} (x - x ^{\prime}) , \label{Green-Eq}
\end{align}
in such a way that the general solution for the 4-potential in the Lorentz gauge is
\begin{align}
A ^{\mu} (x) = \int  G ^{\mu} _{\phantom{\mu} \nu} (x , x ^{\prime}) j ^{\nu} (x ^{\prime}) d ^{4} x .
\end{align}
Since the timelike theory appears to be inconsistent (that is, the theory violates unitary and causality, or both), in this work we specialize to the purely spacelike case $k ^{\mu} \equiv (0 , \textbf{k}) \equiv (0,0,0, \kappa)$. Note that this condition makes the propagation of light anisotropic and defines a class of preferred inertial frames. Without loss of generality, we consider surfaces $\Sigma _{i}$ which are orthogonal to $\textbf{k}$ in which Dirichlet, Neumann or Robin boundary conditions have been imposed. In this way, the GF we consider has translational invariance in the directions $x$ and $y$, while this invariance is broken in the $z$-direction. Exploiting this symmetry we further introduce the Fourier transform in the direction parallel to the surfaces $\Sigma _{i}$, taking the coordinate dependence to be $\textbf{R} = (x - x ^{\prime} , y - y ^{\prime})$, and define
\begin{align}
G ^{\mu} _{\phantom{\mu} \nu} (x,x ^{\prime}) = \int \frac{d ^{2} \textbf{p}}{(2 \pi) ^{2}} e ^{i \textbf{p} \cdot (\textbf{x} - \textbf{x} ^{\prime})} \int \frac{d \omega}{2 \pi} e ^{-i \omega (t-t ^{\prime})} g ^{\mu} _{\phantom{\mu} \nu} (z , z ^{\prime}) , \label{GreenFunc}
\end{align}
where $\textbf{p} = (p _{x} , p _{y})$ is the momentum parallel to $\Sigma _{i}$. In Eq. (\ref{GreenFunc}) we have suppressed the dependence of the reduced GF $g ^{\mu} _{\phantom{\mu} \nu}$ on $\omega$ and $\textbf{p}$.

The substitution of Eq. (\ref{GreenFunc}) into Eq. (\ref{Green-Eq}) yields the reduced GF equation
\begin{align}
\left( \tilde{\Box} \eta ^{\mu} _{\phantom{\mu} \nu} - 2 i \kappa \epsilon ^{3 \mu \beta} _{\phantom{3 \mu \beta} \nu} p _{\beta} \right) g ^{\nu} _{\phantom{\nu} \alpha} (z , z ^{\prime}) = \eta ^{\mu} _{\phantom{\mu} \alpha} \delta (z - z ^{\prime}) , \label{RedGreenEq}
\end{align}
where $p ^{\alpha} = (\omega , \textbf{p} , 0)$ and $\tilde{\Box} = \textbf{p} ^{2} - \omega ^{2} - \partial ^{2} _{z}$. We now must solve the reduced GF equation for the various components. At this point it is worth mentioning that the authors in Ref. \cite{Mariana} say that they derive the GF for a Chern-Simons like theory; however, they used the wrong equation of motion $[ - \partial ^{2} + (k _{AF}) ^{2} ] \epsilon ^{\mu \nu \alpha \beta} \partial _{\alpha} A _{\beta} = 0 $, which leads to a simple analysis in terms of a massive scalar field. Thus its solution is appropriate for the Proca theory, rather than for the Chern-Simons theory. Indeed, in Ref. \cite{Milton}, Milton presents a detailed derivation of the Casimir force for massive photons using different methods. In this section we derive the correct Green's function which solves the equations of motion for the (3+1)D Chern-Simons theory.

The solution to Eq. (\ref{RedGreenEq}) is simple but not straightforward. Since the coefficient $\kappa$ is assumed to be small, to solve it we employ a method similar to that used for obtaining the scattering amplitudes in quantum mechanics, in which the Schr\"{o}dinger equation can be written as an integral equation, the Lippmann-Schwinger equation, which can be iterated to obtain the Born series. Indeed, the Lippmann-Schwinger equation for Green's operator is called the resolvent identity. In the problem at hand let us consider that the free (with $\kappa = 0$) reduced GF is known, being the solution of $\tilde{\Box} \mathfrak{g} (z , z ^{\prime}) = \delta ( z - z ^{\prime})$ in the region $\mathfrak{D} \subseteq \mathbb{R}$ and satisfying appropriate boundary conditions on the surfaces $\Sigma _{i} \subseteq \mathbb{R} ^{2} = \left\lbrace (x,y) : x,y \in \mathbb{R} \right\rbrace $.

Now Eq. (\ref{RedGreenEq}) can be directly integrated using the free reduced GF. We thus establish the integral equation
\begin{align}
g ^{\mu} _{\phantom{\mu} \nu} (z , z ^{\prime}) &= \eta ^{\mu} _{\phantom{\mu} \nu} \mathfrak{g} (z , z ^{\prime}) + 2 i \kappa \epsilon ^{3 \mu \alpha} _{\phantom{3 \mu \alpha} \beta} p _{\alpha}  \int _{\mathfrak{D}} \mathfrak{g} (z , z ^{\prime \prime}) g ^{\beta} _{\phantom{\beta} \nu} (z ^{\prime \prime} , x ^{\prime}) dz ^{\prime \prime} . \label{RedGreenEq2}
\end{align}
Suppose we take this expression for $g ^{\beta} _{\phantom{\beta} \nu}$, and plug it under the integral sign. Iterating this procedure, we obtain a formal series for $g ^{\mu} _{\phantom{\mu} \nu}$. At leading-order in the LV coefficient $\kappa$, the reduced GF can be written as the sum of two terms,
\begin{align}
g ^{\mu} _{\phantom{\mu} \nu} (z,z ^{\prime}) = \eta ^{\mu} _{\phantom{\mu} \nu} \mathfrak{g} (z,z ^{\prime}) + \textbf{g} ^{\mu} _{\phantom{\mu} \nu} (z,z ^{\prime}).
\end{align}
The first term provides the propagation in the absence of Lorentz violation, while the second term, which can be shown to be
\begin{align}
\textbf{g} ^{\mu} _{\phantom{\mu} \nu} (z,z ^{\prime}) &=  2 i \kappa \epsilon ^{\mu \phantom{\nu} \alpha 3} _{\phantom{\mu} \nu} p _{\alpha} E (z,z ^{\prime}) - 4 \kappa ^{2} \left[ p ^{\mu} p _{\nu} - (\eta ^{\mu} _{\phantom{\mu} \nu} + n ^{\mu} n _{\nu}) p ^{2} \right] M (z,z ^{\prime}) , \label{Sol-RGF}
\end{align}
encodes the Lorentz symmetry breakdown. In deriving Eq. (\ref{Sol-RGF}) we have used the identity  $\epsilon ^{3 \mu \beta} _{\phantom{3 \mu \beta} \alpha} \epsilon ^{3 \alpha \gamma} _{\phantom{3 \alpha \gamma} \nu} p _{\beta} p _{\gamma} = p ^{\mu} p _{\nu} - (\eta ^{\mu} _{\phantom{\mu} \nu} + n ^{\mu} n _{\nu}) p ^{2}$, and
\begin{align}
E (z,z ^{\prime}) &= \int _{\mathfrak{D}} \mathfrak{g} (z,z ^{\prime \prime}) \mathfrak{g} ( z ^{\prime \prime} , z ^{\prime} ) dz ^{\prime \prime} , \label{E-function} \\ M (z,z ^{\prime}) &= \int _{\mathfrak{D}} \mathfrak{g} (z,z ^{\prime \prime}) E (z ^{\prime \prime} , z ^{\prime}) dz ^{\prime \prime} . \label{M-function} 
\end{align}
Here $n ^{\mu} = (0,0,0,1)$ is the normal to the surfaces $\Sigma _{i}$. In \cref{App} we present the evaluation of these functions for three simple cases: for a) free-space, b) parallel conducting plates and c) an infinite conducting plate.

Clearly, the full GF matrix $G ^{\mu} _{\phantom{\mu} \nu}$ can also be written as the sum of two terms,
\begin{align}
G ^{\mu} _{\phantom{\mu} \nu} (x,x ^{\prime}) = \eta ^{\mu} _{\phantom{\mu} \nu} \mathcal{G} (x,x ^{\prime}) + \textbf{G} ^{\mu} _{\phantom{\mu} \nu} (x,x ^{\prime}) , \label{Green-Split}
\end{align}
where $\mathcal{G}$ and $\textbf{G} ^{\mu} _{\phantom{\mu} \nu}$ are the Fourier transformations of $\mathfrak{g} (z,z ^{\prime})$ and $\textbf{g} ^{\mu} _{\phantom{\mu} \nu}$, respectively, as defined in Eq. (\ref{GreenFunc}). It is worth mentioning that the second term satisfies the Lorentz gauge condition, i.e. $\partial _{\mu} \textbf{G} ^{\mu} _{\phantom{\mu} \nu} = 0$. The proof follows from the reduced GF: $\partial _{\mu} \textbf{G} ^{\mu} _{\phantom{\mu} \nu} \propto \int p _{\mu} \textbf{g} ^{\mu} _{\phantom{\mu} \nu}$, which vanishes given that $\epsilon ^{\mu \phantom{\nu} \alpha 3} _{\phantom{\mu} \nu} p _{\mu} p _{\alpha} = 0$ and $p _{\mu} n ^{\mu} = 0$. 

The reciprocity between the position of the unit charge and the position at which the GF is evaluated, $G _{\mu \nu} (x,x ^{\prime}) = G _{\nu \mu} (x ^{\prime} , x)$, is one of its most remarkable properties. From Eq. (\ref{GreenFunc}), this condition requires
\begin{align}
g _{\mu \nu} (z,z ^{\prime} , p ^{\alpha}) = g _{\nu \mu} (z ^{\prime} , z , - p ^{\alpha}) ,
\end{align}
which we verify directly from Eq. (\ref{Sol-RGF}).

\section{Vacuum stress-energy tensor} \label{VS-section}

In section \ref{model} we derived the stress-energy tensor (SET) for this theory and we showed that it can be written as the sum of two terms:
\begin{align}
\Theta ^{\mu \nu} = T ^{\mu \nu} + \Xi ^{\mu \nu} . \label{SET}
\end{align}
The first term,
\begin{align}
T ^{\mu \nu} = - F ^{\mu \alpha} F ^{\nu} _{\phantom{\nu} \alpha} + \frac{1}{4} \eta ^{\mu \nu} F ^{\alpha \beta} F _{\alpha \beta} ,
\end{align}
is the standard Maxwell stress-energy tensor, while the second,
\begin{align}
\Xi ^{\mu \nu} = - k ^{\nu} \tilde{F} ^{\mu \alpha} A _{\alpha} ,
\end{align}
explicitly depends on the LV coefficients $k ^{\mu}$. Now we address the problem of the vacuum expectation value of the SET, to which we will refer simply as the vacuum stress (VS).

The local approach to compute the VS was initiated by Brown and Maclay who calculated the renormalized stress tensor by means of GF techniques. Therein, the VS can be obtained from appropriate derivatives of the GF, in virtue of the formula
\begin{equation}
G ^{\mu \nu} \left( x , x ^{\prime} \right) = - i \left< 0 \right| \hat{\mathcal{T}}  A ^{\mu} \left( x \right) A ^{\nu} \left( x ^{\prime} \right)  \left| 0 \right> . \label{VacuumGreen}
\end{equation}
Using the standard point splitting technique and taking the vacuum expectation value of the SET we find
\begin{align}
\left<  \Theta ^{\mu \nu} \right> = \left<  T ^{\mu \nu} \right> + \left<  \Xi ^{\mu \nu} \right> , \label{VS}
\end{align}
where the first term,
\begin{align}
\left<  T ^{\mu \nu} \right> = i \lim _{x ^{\prime} \rightarrow x} \Big[ - \partial ^{\mu}  \partial ^{\prime \nu} G ^{\lambda} _{\phantom{\lambda} \lambda} + \partial ^{\mu} \partial _{\lambda} ^{\prime} G ^{\lambda \nu} + \partial ^{\lambda} \partial ^{\prime \nu} G ^{\mu} _{\phantom{\mu} \lambda} - \partial ^{\prime \lambda} \partial _{\lambda} G ^{\mu \nu} + \frac{1}{2} \eta ^{\mu \nu} \left( \partial ^{\alpha} \partial _{\alpha} ^{\prime} G ^{\lambda} _{\phantom{\lambda} \lambda} - \partial ^{\alpha} \partial _{\beta} ^{\prime} G ^{\beta} _{\phantom{\beta} \alpha} \right) \Big] \label{Maxwell-VS}
\end{align}
is the VS of the standard Maxwell SET, and
\begin{align}
\left<  \Xi ^{\mu \nu} \right> = - 2 i k ^{\nu} \epsilon ^{\mu \alpha \beta \gamma } \lim _{x ^{\prime} \rightarrow x} \partial _{\beta} ^{\prime} G _{\gamma \alpha} .
\end{align}
Here we have omitted the dependence of $G ^{\mu \nu}$ on $x$ and $x ^{\prime}$. This result can be further simplified as follows. Since the GF can be written as the sum of two terms, then Eq. (\ref{Maxwell-VS}) can also be written in the same way, i.e.
\begin{align}
\left< T ^{\mu \nu} \right> = \left< t ^{\mu \nu} \right> + \left< \textbf{T} ^{\mu \nu} \right> .
\end{align}
The first term,
\begin{align}
\left< t ^{\mu \nu} \right> = - i \lim _{x ^{\prime} \rightarrow x} \left( 2 \partial ^{\mu} \partial ^{\prime \nu} - \frac{1}{2} \eta ^{\mu \nu} \partial ^{\lambda} \partial _{\lambda} ^{\prime} \right) \mathcal{G} \left( x , x ^{\prime} \right) , \label{SE-free}
\end{align}
is the vacuum stress in the absence of Lorentz violation. In obtaining Eq. (\ref{SE-free}) we used the fact that the zeroth-order GF (with $\kappa = 0 $) is diagonal, i.e. it is equal to $\eta ^{\mu} _{\phantom{\mu} \nu} \mathcal{G} (x,x^{\prime})$. The second term can be simplified since the GF $\textbf{G} ^{\mu} _{\phantom{\mu} \nu}$ satisfies the Lorentz gauge condition $\partial _{\mu} \textbf{G} ^{\mu} _{\phantom{\mu} \nu} = 0$. In this way,
\begin{equation}
\left< \textbf{T} ^{\mu \nu} \right> = - i \lim _{x ^{\prime} \rightarrow x} \left[ \partial ^{\mu} \partial ^{\prime \nu} \textbf{G} + \partial ^{\prime \lambda} \partial _{\lambda} \left( \textbf{G} ^{\mu \nu} - \frac{1}{2} \eta ^{\mu \nu} \textbf{G} \right) \right] , \label{VS-2}
\end{equation}
where $\textbf{G} = \textbf{G} ^{\mu} _{\phantom{\mu} \mu}$ is the trace of $\textbf{G} ^{\mu} _{\phantom{\mu} \nu}$. Finally the last term in Eq. (\ref{VS}) can be written as
\begin{align}
\left<  \Xi ^{\mu \nu} \right> = - 2 i k ^{\nu} \epsilon ^{\mu \alpha \beta \gamma } \lim _{x ^{\prime} \rightarrow x} \partial _{\beta} ^{\prime} \textbf{G} _{\gamma \alpha} , \label{VS-3}
\end{align}
where we used again that the zeroth-order GF is diagonal. 

\section{Casimir effect} \label{CasEff-Section}

Now let us consider the problem of calculating the renormalized VS $\left<  \Theta ^{\mu \nu} \right> _{\mathrm{ren}}$, which is obtained as the difference between the VS in the presence of boundaries and that of the vacuum. For two parallel conducting plates separated by a distance $D$ in the $z$ direction, one can construct the renormalized expectation value of the stress-energy tensor, using conservation, tracelessness and symmetry arguments. The result is that it is uniform between the plates:
\begin{align}
\left< t ^{\mu \nu} \right> _{\mathrm{ren}} =  - \frac{\pi ^{2}}{720 D ^{4}} \left( \eta ^{\mu \nu} + 4 n ^{\mu} n ^{\nu} \right) , \label{BM-T}
\end{align}
where $n ^{\mu} = (0,0,0,1)$ is the unit normal to the plates. The Casimir stress is obtained by differentiating the Casimir energy $E _{C} = \left< t ^{00} \right> _{\mathrm{ren}} D$ with respect to $D$, i.e. $F _{C} = - d E _{C} / dD = - \pi ^{2} /240 D ^{4}$.

We turn now with the vacuum stress between two perfectly, conducting, infinite plates, separated by a distance $D$, embedded in the infinite Lorentz-violating vacuum. We orient the coordinate frame so that the plates are perpendicular to the background LV vector $k ^{\mu} = (0, \textbf{k}) = \kappa n ^{\mu}$. 

Using the Fourier representation of the GF in Eq. (\ref{GreenFunc}), together with the symmetry of the problem and the reduced Green's function given by Eq. (\ref{Sol-RGF}), Eq. (\ref{VS-2}) can be written as
\begin{align}
\left< \textbf{T} ^{\mu \nu} \right> &= - 4 i \kappa ^{2} \int \frac{d ^{2}\textbf{p}}{(2 \pi) ^{2}} \int \frac{d \omega}{2 \pi} \left( p ^{\mu} p ^{\nu} + n ^{\mu} n ^{\nu} p ^{2} \right) \lim _{z ^{\prime} \rightarrow z } \left( p ^{2} + \partial _{z} \partial _{z} ^{\prime} \right) M (z,z ^{\prime}) . \label{VS-4}
\end{align}
From the rotational invariance around the $z$ axis, the components of the stress perpendicular to $n ^{\mu}$, $\left< \Theta ^{11} \right>$ and $\left< \Theta ^{22} \right>$, are equal. In addition, from the mathematical structure of Eq. (\ref{VS-3}) we find the relation $\left< \textbf{T} ^{00} \right> = - \left< \textbf{T} ^{11} \right>$. These results, together with the tracelessness of $\left< \textbf{T} ^{\mu \nu} \right>$, allow us to write the VS of Eq. (\ref{VS-4}) in the form
\begin{align}
\left< \textbf{T} ^{\mu \nu} \right> &= \left( \eta ^{\mu \nu} + 4 n ^{\mu} n ^{\nu} \right) f (\kappa , z) ,  \label{VS-5} 
\end{align}
where
\begin{align}
f (\kappa , z) = \frac{4 \kappa ^{2}}{i} \int \frac{d ^{2}\textbf{p}}{(2 \pi) ^{2}} \int \frac{d \omega}{2 \pi} \omega ^{2} \lim _{z ^{\prime} \rightarrow z } \left( p ^{2} + \partial _{z} \partial _{z} ^{\prime} \right) M (z,z ^{\prime}) . \label{f-func}
\end{align}
Note that this term exhibits the same tensor structure as the result obtained by Brown and Maclay, but we obtain a $z$-dependent VS due to Lorentz violation. Using similar arguments, the last contribution to the VS given by Eq. (\ref{VS-3}) can be written as
\begin{align}
\left< \Xi ^{\mu \nu} \right> &= n ^{\mu} n ^{\nu} g (\kappa , z ) , \label{VS-6} 
\end{align}
where
\begin{align}
g (\kappa , z ) = - 8 i \kappa ^{2} \int \frac{d ^{2}\textbf{p}}{(2 \pi) ^{2}} \int \frac{d \omega}{2 \pi} p ^{2} \lim _{z ^{\prime} \rightarrow z} E (z,z ^{\prime}) . \label{g-func}
\end{align}
Therefore the renormalized vacuum expectation value of the stress-energy tensor $\Theta ^{\mu \nu}$ between the conducting plates can be written as
\begin{align}
\left< \Theta ^{\mu \nu} \right> _{\mathrm{ren}} &= \left< t ^{\mu \nu} \right> _{\mathrm{ren}} + \left( \eta ^{\mu \nu} + 4 n ^{\mu} n ^{\nu} \right) f _{\mathrm{ren}} (\kappa , z) + n ^{\mu} n ^{\nu} g _{\mathrm{ren}} (\kappa , z ) , \label{VS-fin}
\end{align}
where $\left< t ^{\mu \nu} \right> _{\mathrm{ren}}$, which is given by Eq. (\ref{BM-T}), is the result obtained by Brown and Maclay in Lorentz symmetric electrodynamics. Here, the renormalized functions $f _{\mathrm{ren}} (\kappa , z)$ and $g _{\mathrm{ren}} (\kappa , z)$ are defined as the the difference between the functions in the presence of the plates and that of the vacuum. Note that the vacuum stress is symmetric, even when the SET is non-symmetric. Next we derive the Casimir force between the plates.

\subsection{Global Casimir energy density} \label{CasEnergySec}

The Casimir energy is defined as the energy per unit area stored in the electromagnetic field between the plates, i.e.
\begin{align}
\mathcal{E} _{C} = \int _{0} ^{D} \left< \Theta ^{00} \right> _{\mathrm{ren}} dz = - \frac{\pi ^{2}}{720 D ^{3}} + E _{\mathrm{ren}} (\kappa) , \label{CasEnergy}
\end{align}
where we have defined the function
\begin{align}
E _{\mathrm{ren}} (\kappa) = \int _{0} ^{D} f _{\mathrm{ren}} (\kappa , z) dz . \label{I-ren}
\end{align}
The first term in Eq. (\ref{CasEnergy}) is the usual Casimir energy density between two parallel conducting plates in Lorentz symmetric electrodynamics. Now we must evaluate the function $I _{\mathrm{ren}} (\kappa)$. First we consider the VS in the presence of the plates and define the function $E _{\parallel} (\kappa) \equiv \int _{0} ^{D} f _{\parallel} (\kappa , z) dz $. Using the function $M _{\parallel} (z,z ^{\prime})$ of Eq. (\ref{M-Plates-FIN}) we obtain:
\begin{align}
E _{\parallel} (\kappa) &= - 4 i \kappa ^{2} \int \frac{d ^{2} \textbf{p}}{(2 \pi) ^{2}} \int \frac{d \omega}{2 \pi} \frac{\omega ^{2} }{8 p ^{4}} \Big[ 4 - p D \cot (p D) - p ^{2} D ^{2} \left[ 1 + 2 p D \cot (p D) \right] \csc ^{2} (p D) \Big] . \label{I-plates}
\end{align}
Notice that the first term in the integrand corresponds to a constant energy density, independent of $D$, so it may be discarded as irrelevant. The resulting integral can be evaluated as follows. We first write the momentum element as $d ^{2} \textbf{p} = \vert \textbf{p} \vert d \vert \textbf{p} \vert d \theta$ and integrate $\theta$ from $0$ to $2 \pi$. Next we perform a Wick rotation such that $\omega \rightarrow i \zeta$, then replace $\zeta$ and $\vert \textbf{p} \vert$ by the plane polar coordinates $\zeta = \frac{\xi }{D} \cos \varphi$, $\vert \textbf{p} \vert = \frac{\xi }{D} \sin \varphi$ and finally integrate $\varphi$ from $0$ to $\pi / 2$. In this way, Eq. (\ref{I-plates}) becomes
\begin{align}
E _{\parallel} (\kappa) = \frac{\kappa ^{2}}{24 \pi ^{2} D} \int _{0} ^{\infty} \xi \left[ \coth \xi + \xi \left( 1 + 2 \xi \coth \xi \right) \mbox{csch} ^{2} \xi \right] d \xi . \label{I-plates2}
\end{align}
We can treat this result as containing two pieces. The first integral, which involves $\xi \coth \xi$, is divergent, so that we will retain such. The remaining part is convergent and it can be evaluated. The result is
\begin{align}
E _{\parallel} (\kappa) = \frac{\kappa ^{2}}{24 \pi ^{2} D} \left[ \frac{2 \pi ^{2}}{3} + \int _{0} ^{\infty} \xi \coth \xi d \xi \right] . \label{I-plates3}
\end{align}
Now we must evaluate the analogue function in vacuum. Using the function $M _{0} (z,z ^{\prime})$ of Eq. (\ref{M0-FreeSpace}) we obtain
\begin{align}
E _{0} (\kappa) = 4  \kappa ^{2} \int \frac{d ^{2} \textbf{p}}{(2 \pi) ^{2}} \int \frac{d \omega}{2 \pi} \frac{\omega ^{2} D}{8 p ^{3}} . \label{I-vac}
\end{align}
The integral can be treated similarly. Integrating the angle in momentum space, performing a Wick rotation and introducing the plane polar coordinates defined before we find:
\begin{align}
E _{0} (\kappa) = \frac{\kappa ^{2}}{24 \pi ^{2} D} \int _{0} ^{\infty} \xi d \xi , \label{I-vac2}
\end{align}
which clearly is a divergent integral. Subtracting both contributions to obtain the renormalized function, i.e. $E _{\mathrm{ren}} (\kappa) = E _{\parallel} (\kappa) - E _{0} (\kappa)$, we obtain
\begin{align}
E _{\mathrm{ren}} (\kappa) = \frac{\kappa ^{2}}{24 \pi ^{2} D} \left[ \frac{2 \pi ^{2}}{3} + \int _{0} ^{\infty} \xi  \left(\coth \xi - 1 \right) d \xi  \right] , \label{I-ren2}
\end{align}
and the resulting integral is perfectly convergent, with the result $\pi ^{2} / 12$. Therefore the final expression for the renormalized function is $E _{\mathrm{ren}} (\kappa) = \kappa ^{2} / 32 D$, and thus the Casimir energy becomes
\begin{align}
\mathcal{E} _{C} =  - \frac{\pi ^{2}}{720 D ^{3}} + \frac{\kappa ^{2} }{32 D} . \label{CasEnergy2}
\end{align}
The Casimir stress is obtained by differentiating the Casimir energy with respect to $D$, i.e.
\begin{align}
\mathcal{F} _{C} = - \frac{d \mathcal{E} _{C}}{d D} =  - \frac{\pi ^{2}}{240 D ^{4}} + \frac{\kappa ^{2} }{32 D ^{2}} . \label{CasForce}
\end{align}

\subsection{Stress on the plates}

Now let us derive the same result by using the normal-normal component of the vacuum stress energy-tensor, i.e. $\left< \Theta ^{zz} \right> $. As suggested by the structure of the vacuum stress (\ref{VS-fin}), the force per unit area can be written as the sum of two terms,
\begin{align}
\mathcal{F} _{C} = - \frac{\pi ^{2}}{240 D ^{4}} + F _{\mathrm{ren}} (\kappa) ,
\end{align}
where the first term is the Casimir stress in the absence of Lorentz violation, and the function $F _{\mathrm{ren}} (\kappa) = \left< \textbf{T} ^{zz} + \Xi ^{zz} \right> _{\mathrm{ren}} $ encodes the renormalized part to the Casimir stress due to the Lorentz-violating coefficient. 

By virtue of the boundary conditions upon the plates, we compute the normal-normal component of the vacuum stress tensor on the boundaries as
\begin{align}
\left< \textbf{T} ^{zz} \right> = - 12 i \kappa ^{2} \int \frac{d ^{2}\textbf{p}}{(2 \pi) ^{2}} \int \frac{d \omega}{2 \pi} \omega ^{2} \lim _{z ^{\prime} \rightarrow z = 0,D } \partial _{z} \partial _{z} ^{\prime} M (z,z ^{\prime}) , \label{Tzz-plates}
\end{align}
since $M (z,z ^{\prime}) = 0$ at $z=0$ and $z=D$. First we consider the contribution due to the electromagnetic field confined between the plates. The relevant function $M _{\parallel} (z,z ^{\prime})$ is given by Eq. (\ref{M-Plates-FIN}). The resulting integral can be treated in the same way as that of Eq. (\ref{I-ren}). Integrating the angle in momentum space, performing a Wick rotation and introducing the plane polar coordinates, we find
\begin{align}
\left< \textbf{T} ^{zz} \right> _{\parallel} &= - \frac{\kappa ^{2}}{32 \pi ^{2} D ^{2}} \int _{0} ^{\infty} \xi \mbox{csch} ^{3} \xi \Big[ \cosh (3 \xi) - (1+8 \xi ^{2}) \cosh \xi + 4 \xi \sinh \xi \Big] d \xi . 
\end{align}
This integral does not exist. However, as in the standard case, it can be appropriately regularized. To do this, let us consider the normal-normal component of the vacuum stress due to the electromagnetic field at the right side of the plate at $z=D$. Using the appropriate reduced GF $\mathfrak{g} _{\mid} (z,z ^{\prime})$ which vanishes at $z = D$ and has outgoing boundary conditions as $z \rightarrow \infty$, i.e. $\mathfrak{g} _{\mid} \sim e ^{i p z}$, one can construct the associated $M _{\mid} (z,z ^{\prime})$-function. The calculation is presented in \cref{App}. The relevant function is given by Eq. (\ref{M-1Plate}). The corresponding normal-normal component of the vacuum stress at $z = D$ is then
\begin{align}
\left< \textbf{T} ^{zz} \right> _{\mid} = - \frac{\kappa ^{2}}{8 \pi ^{2} D ^{2}} \int _{0} ^{\infty} \xi d \xi , \label{Tzz-vacuum}
\end{align}
which also diverges. So, from the discontinuity in $\left< \textbf{T} ^{zz} \right>$, that is, the difference $ \left< \textbf{T} ^{zz} \right> _{\parallel} - \left< \textbf{T} ^{zz} \right> _{\mid} $, we find the Lorentz-violating contribution to the force per unit area on the plate:
\begin{align}
F _{\mathrm{ren}} (\kappa) &= - \frac{\kappa ^{2}}{32 \pi ^{2} D ^{2}} \int _{0} ^{\infty} \xi \Big\{ \mbox{csch} ^{3} \xi \big[ \cosh (3 \xi) - (1+8 \xi ^{2}) \cosh \xi + 4 \xi \sinh \xi \big] - 4 \Big\} d \xi .
\end{align}
The resulting integral is perfectly convergent, with the result $- \pi ^{2}$. In this way, the stress on the plate is $\mathcal{F} _{C} = - \frac{\pi ^{2}}{240 D ^{4}} + \frac{\kappa ^{2}}{32 D ^{2}}$, which is the same as obtained in the previous subsection.

\subsection{Local effects}

Heretofore, we have considered the global Casimir effect: the total energy of a field configuration or the force per unit area on a bounding surface \cite{Milton}. Local properties of the quantum vacuum induced by the presence of boundaries are of broad interest in quantum field theory \cite{Candelas} and they must be understood if one is to correctly interpret the inherent divergences in the theory. 

\begin{figure}
\includegraphics[width = 8 cm]{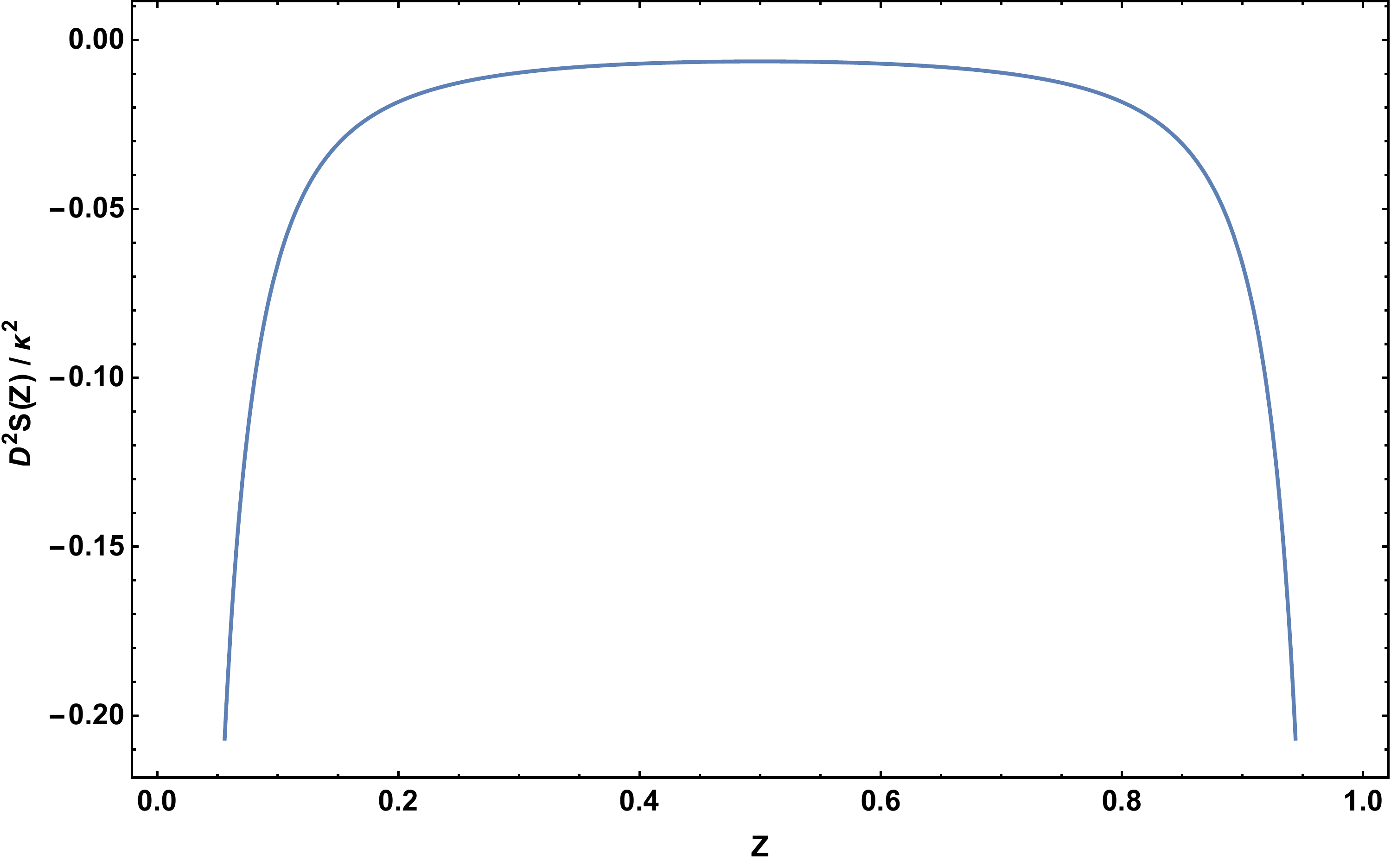}
\caption{\small The singular part of the Lorentz-violating correction to the local energy density between two parallel plates.}
\label{LocalEnergyPlot}
\end{figure}

The local energy density in Lorentz-symmetric electrodynamics has been discussed extensively in the literature \cite{Milton, Sopova}; however, the local effects in Lorentz-violating theories have not been considered. Here we aim to fill in this gap. We begin the analysis by considering an electromagnetic field confined between two parallel conducting plates at $z = 0$ and $z = D$, for which the energy density per unit volume between the plates is
\begin{align}
\left< \Theta ^{00} \right> (z) &= \left< t ^{00} \right>  (z) + \left< \textbf{T} ^{00} \right> (z) . \label{T00-local}
\end{align}
A detailed analysis of the local effects due to the first term (in the absence of Lorentz violation) is presented in Ref. \cite{Milton}. Here we concentrate on the Lorentz-violating contribution, 
\begin{align}
\left< \textbf{T} ^{00} \right> (z)  = \frac{4 \kappa ^{2}}{i} \int \frac{d ^{2} \textbf{p}}{( 2 \pi ) ^{2}} \int \frac{d \omega}{2 \pi} \omega ^{2} \lim _{z ^{\prime} \rightarrow z} \left( p ^{2} + \partial _{z} \partial _{z} ^{\prime} \right) M _{\parallel} (z,z ^{\prime}) , \label{T00-local2}
\end{align}
where $M _{\parallel} (z,z ^{\prime})$ is given by Eq. (\ref{M-Plates-FIN}). This integral can be performed in an analytical fashion. Integrating the angle in momentum space, performing a Wick rotation and introducing the plane polar coordinates, we find
\begin{align}
\left< \textbf{T} ^{00} \right> (z)  = - \frac{\kappa ^{2}}{3 \pi ^{2} D ^{2}} \int _{0} ^{\infty} \lambda ^{4} \left[ F (\lambda) + R (\lambda , Z ) \right]  d \lambda ,
\end{align}
where
\begin{align}
F (\lambda ) &= - \frac{\text{csch} ^3 \lambda}{32 \lambda ^{3}} \Big\{ \left( 8 \lambda ^{2} - 1 \right) \cosh \lambda + \cosh (3 \lambda) + 4 \lambda \sinh \lambda \Big\} , \\ R (\lambda , Z ) &= \frac{\text{csch} ^3 \lambda}{16 \lambda ^{3}} \Big\{ 2 \lambda ( 1 - Z ) \sinh [ \lambda (1 + 2 Z) ] + 2 \lambda (1 - 2 Z ) \sinh [ \lambda (1- 2 Z )] + 2 \lambda Z \sinh [ \lambda ( 3 - 2 Z ) ] \\ & \hspace{5 cm} + \cosh [ \lambda (1 + 2 Z ) ] - 2 \cosh [ \lambda (1- 2 Z) ] + \cosh [ \lambda (3 - 2 Z) ] \Big\} \notag ,
\end{align}
with $Z = z / D$. Integrating we obtain
\begin{align}
\left< \textbf{T} ^{00} \right> (z)  = \frac{\kappa ^{2}}{32 D ^{2}} \left[ 1 - 2 \csc ^{2} (\pi Z) \right]  .
\end{align}
We observe that the $z$-independent term, $\frac{\kappa ^{2}}{32 D ^{2}}$, corresponds to the global renormalized energy density $E  _{\mathrm{ren}} (\kappa)$ obtained in \cref{CasEnergySec}. The second term encodes the local effects, and it can be expressed in terms of the generalized or Hurwitz zeta function as follows:
\begin{align}
\mathcal{S} (Z) \equiv  - \frac{\kappa ^{2}}{16 D ^{2}} \csc ^{2} (\pi Z) = - \frac{\kappa ^{2}}{16 \pi ^{2} D ^{2}} \left[ \zeta (2 , Z) + \zeta (2 , 1 - Z) \right] . \label{DivergentFunc}
\end{align}
This function is plotted in \cref{LocalEnergyPlot}, where we observe that that it diverges quadratically as $z \rightarrow 0 , D$. Its $z$ integral over the region between t$  $he plates diverges linearly. This result reveals a close analogy with the one obtained from the Lorentz-symmetric part. In that case, the singular part depends on $[ \zeta (4 , Z) + \zeta (4 , 1 - Z) ] / D ^{4}$, thus implying that it diverges quartically as $z \rightarrow 0 , D$. The less divergent Lorentz-violating contribution (\ref{DivergentFunc}) can be understood as due to the dimensionfull Chern-Simons coupling $\kappa$. 

To close this section, we show that the badly behaved function (\ref{DivergentFunc}) does not contribute to the force on the plates. Using the integral representation of the Hurwitz zeta function in terms of the Mellin transform,
\begin{equation}
\zeta ( s , q ) = \frac{1}{\Gamma ( s ) } \int _{0} ^{\infty} \frac{ t ^{s-1} e ^{- qt} }{1 - e ^{- t}} dt \quad , \quad \quad \quad \Re [s]>1,\,\,\, \Re [q]>0 ,
\end{equation}
Eq. (\ref{DivergentFunc}) can be expressed as
\begin{align}
\mathcal{S} (Z) = - \frac{\kappa ^{2}}{16 \pi ^{2} } \frac{1}{\Gamma (2)} \int _{0} ^{\infty} \frac{e ^{- z \rho} + e ^{- (D-z) \rho}}{1 - e ^{- \rho D}} \rho d \rho .
\end{align}
If we integrate this term over $z$,
\begin{align}
\int _{0} ^{D} \mathcal{S} (Z) dz = - \frac{\kappa ^{2}}{16 \pi ^{2} } \frac{2}{\Gamma (2)} \int _{0} ^{\infty} d \rho , 
\end{align}
we obtain a divergent constant term ($D$-independent), so it does not contribute to the observable force.

\section{Summary and conclusions} \label{conclusions}

The study of Lorentz violation is actively motivated by the search of quantum gravity effects. At presently attainable energies, such signatures are described by an effective field theory framework called the standard model extension (SME), which contains the standard model, general relativity and all possible operators that break Lorentz symmetry. The Lorentz-violating (LV) coefficients in the SME can arise in various underlying contexts, such as strings, spacetime-foam approaches and non-commutative geometries, to name a few. The value of these coefficients can in principle be measured (or bounded) in experiments. This has allowed a systematic search for a large range of possible Lorentz-violating effects. Indeed, since the predictions of the standard model have been verified experimentally to an extremely high degree of accuracy, a possible route to test quantum gravitational effects is through high-sensitivity measurements of well-known particle physics phenomena, as any deviation from the standard theory is, at least in principle, experimentally testable. 

The main aim of this paper is to analyze the local effects of the quantum vacuum in a particular sector of the minimal SME, namely, the (3+1)D Maxwell-Chern-Simons term. Concretely, we use a local approach to calculate the Casimir force between two parallel conducting plates. This problem was first considered by M. Frank and I. Turan \cite{Mariana} within the SME framework. They calculated the Casimir force between two parallel conducting plates. However, as pointed out by O. G. Kharlanov and V. Ch. Zhukovsky \cite{Kharlanov}, in the aforementioned work the authors used a wrong equation of motion for the electromagnetic potentials, namely, $(- \partial ^{2} + k _{AF} ^{2}) \epsilon ^{\mu \nu \alpha \beta}\partial _{\alpha} A _{\beta} = 0$, instead of the correct equation $( \Box \eta ^{\mu} _{\phantom{\mu} \nu} - 2 k _{\beta} \epsilon ^{\mu \beta \alpha} _{\phantom{\mu \beta \alpha} \nu} \partial _{\alpha} ) A ^{\nu} = 0$. This misinterpreted equation led to an oversimplification of the problem from both the conceptual and theoretical points of view. Conceptually we observe that the wrong equation implies that the photon dispersion relation corresponds to that for a massive photon, similar to what occurs in the (2+1)D Maxwell-Chern-Simons theory. However, in the (3+1)D theory the photon remains massless and the associated dispersion relation is more complicated. Mathematically, the wrong field equation leads to a simple analysis of the problem in terms of a single Green's function; however, the correct field equation suggests that a more subtle indexed (matrix of) Green's function(s) is required due to the magnetoelectric effect (i.e. the mixing between the components of the 4-potential).

The authors in Ref. \cite{Kharlanov} evaluated the Casimir force within this theory using two methods, namely, the zeta function regularization technique and the summation and renormalization of the discrete sum involving the residue theorem. In the present work, we have employed a far superior technique based upon the use of Green's functions. Because the Green's function represents the vacuum expectation value of the time-ordered product of fields, it is possible to compute the vacuum expectation value of the stress-energy tensor, for example, in terms of the Green's function at coincident arguments. We restrict ourselves to the analysis of the (3+1)D Maxwell-Chern-Simons theory with a spacelike LV coefficient and we construct the relevant (leading-order) Green's function which satisfies given boundary conditions. Using these results, we evaluate the renormalized vacuum expectation value of the stress-energy tensor and we compute the Casimir force between two parallel conducting plates separated by a distance $D$, with the result $\mathcal{F} _{C} = - \frac{\pi ^{2}}{240 D ^{4}} + \frac{\kappa ^{2}}{32 D ^{2}}$. We recognize the first term as the Casimir force in the absence of Lorentz violation, while the $\kappa$-dependent term is due to such LV coefficient, which has a different functional dependence on the distance between the plates ($1/D ^{2}$). This result is expected from dimensional analysis for the second order LV contributions. Due to the limited precision of the present experimental measurements of the Casimir force between parallel plates (15 \% precision in the $0.5-3 \mu$m range), no useful bounds on the LV coefficients can be obtained from the results in this work. We also analyze the behavior of the local energy density when approaching the plates, which is found to be quadratically divergent according to $\zeta (2, z/D) + \zeta (z , 1-z/D)$, where $\zeta (s,q)$ is the Hurwitz zeta function. In the standard case is obtained a quartically divergent term. The less divergent Lorentz-violating contribution can be understood as due to the dimensionfull Chern-Simons coupling $\kappa$. Using the integral representation of the Hurwitz zeta function in terms of the Mellin transform, we demonstrate tha such term does not contribute to the measurable macroscopic force.

The present work can be further generalized in a variety of ways. For example, the Green's function for different geometries can also be constructed using the same perturbative procedure. On the other hand, our analysis can also be applied to ponderable media. More precisely, we can consider a semi-infinite planar material medium with dielectric constant $\varepsilon$ for which the reduced Green's function $\mathfrak{g} _{\varepsilon} (z,z ^{\prime})$ is known. Now we can use it to evaluate the associated $E _{\varepsilon} (z,z ^{\prime})$ and $M _{\varepsilon} (z,z ^{\prime})$ functions, which are what we require to study the Lorentz-violating effects. This scenario could be more useful to establish bounds on the SME coefficients, since in this case first order LV contributions could appear.

\appendix

\section{Evaluation of the functions $E (z,z ^{\prime})$ and $M (z,z ^{\prime})$} \label{App}

Here we shall evaluate the functions $E (z,z ^{\prime})$ and $M (z,z ^{\prime})$, defined by Eqs. (\ref{E-function}) and (\ref{M-function}), respectively. To do this, we use the reciprocity symmetry of the reduced GF, $\mathfrak{g} (z,z ^{\prime}) = \mathfrak{g} (z ^{\prime} , z )$, which implies the same property for the functions $E (z,z ^{\prime})$ and $M (z,z ^{\prime})$, i.e.
\begin{align}
E (z , z ^{\prime}) = E (z ^{\prime} , z) \qquad , \qquad M (z , z ^{\prime}) = M (z ^{\prime} , z) .
\end{align}
In this way, we can compute the required integrals for a particular case, for example $z < z ^{\prime}$, and then generalize the result by replacing $z \rightarrow z _{<}$ and $z ^{\prime} \rightarrow z _{<}$, where $z _{>}$ ($z _{<}$) will be the greater (lesser) between $z$ and $z ^{\prime}$.

Let $\mathcal{D} = [a,b]$ be the domain in which the reduced GF is defined. Assuming that $z < z ^{\prime}$, the domain of integration can be expressed as $\mathcal{D} = [a,z] + [z,z ^{\prime}] + [z ^{\prime},b]$. The function $E (z , z ^{\prime}) $ can thus be written as
\begin{align}
E (z , z ^{\prime}) = \left( \int _{a} ^{z} + \int _{z} ^{z ^{\prime}} + \int _{z ^{\prime}} ^{b}  \right) \mathfrak{g} (z , z ^{\prime \prime})  \mathfrak{g} (z ^{\prime \prime} , z ^{\prime}) dz ^{\prime \prime} , \label{E-func-App}
\end{align}
and the function $M (z , z ^{\prime})$ takes the form
\begin{align}
M (z , z ^{\prime}) = \left( \int _{a} ^{z} + \int _{z} ^{z ^{\prime}} + \int _{z ^{\prime}} ^{b}  \right) \mathfrak{g} (z , z ^{\prime \prime})  E (z ^{\prime \prime} , z ^{\prime}) dz ^{\prime \prime} . \label{M-func-App}
\end{align}
Now we compute these integrals for some cases of particular interest in this paper. 

\subsection*{Free-space}

First we consider the free-space reduced Green's function
\begin{align}
\mathfrak{g} _{0} (z,z ^{\prime}) = \frac{i}{2 p } e ^{ip ( z _{>} - z _{<} )} , \label{g0-FreeSpace}
\end{align}
which is defined on the real line, i.e. $\mathcal{D} = \mathbb{R}$. The associated $E _{0}$-function can thus be written as

\begin{widetext}
\[
E _{0} (z,z ^{\prime}) = - \frac{1}{4 p ^{2}} \left[ \int _{- \infty} ^{z} e ^{i p ( z - z ^{\prime \prime} )} e ^{i p ( z ^{\prime} - z ^{\prime \prime})} dz ^{\prime \prime} + \int _{z} ^{z ^{\prime}} e ^{i p ( z ^{\prime \prime} - z)} e ^{i p ( z ^{\prime} - z ^{\prime \prime})} dz ^{\prime \prime} + \int _{z ^{\prime}} ^{\infty} e ^{i p ( z ^{\prime \prime} -z )} e ^{i p ( z ^{\prime \prime} - z ^{\prime})} dz ^{\prime \prime} \right]  \label{E0}
\]
\end{widetext}

The result for this particular case ($z<z ^{\prime}$) can be obtained in a simple fashion. The general result takes the form
\begin{align}
E _{0} (z,z ^{\prime}) = - \frac{i + p (z _{>} - z _{<})}{4 p ^{3}} e^{i p (z _{>} - z _{<})} . \label{E0-FreeSpace}
\end{align}
Now, using the reduced GF of Eq. (\ref{g0-FreeSpace}) and the previous result (\ref{E0-FreeSpace}), we can compute the associated $M _{0}$-function according to Eq. (\ref{M-func-App}). The result is
\begin{align}
M _{0} (z,z ^{\prime}) = \frac{p ^{2} (z _{>} - z _{<}) ^{2}+ 3 i p (  z _{>} - z _{<}) - 3}{16 i p ^{5}} e^{i p (z _{>} - z _{<})} . \label{M0-FreeSpace}
\end{align}
Performing a Wick rotation one can further see that these functions satisfy the required boundary conditions at infinity, i.e. $E _{0} (z \rightarrow \infty ,z ^{\prime}) , M _{0} (z \rightarrow \infty , z ^{\prime}) \rightarrow 0$.

\subsection*{Parallel conducting plates}

Now let us consider the case of two parallel conducting plates separated by a distance $D$. The relevant reduced GF for this configuration is
\begin{align}
\mathfrak{g} _{\parallel} (z,z ^{\prime}) = \frac{\sin [p z _{<}] \sin [p (D - z _{>})]}{p \sin [pD]} , \label{RedGFPlates}
\end{align}
which is defined on the domain $\mathcal{D} _{\parallel} = [0,D]$ and satisfies the boundary conditions $\mathfrak{g} _{\parallel} (0,z ^{\prime}) = \mathfrak{g} _{\parallel} (D,z ^{\prime}) = 0$. The associated $E _{\parallel}$-function can be written as
\begin{align}
E _{\parallel} (z,z ^{\prime}) = \frac{1}{p ^{2} \sin ^{2} [pD]} & \left[ \int _{0} ^{z} \sin ^{2} [pz ^{\prime \prime}] \sin [p (D - z)] \sin [p (D - z ^{\prime})] dz ^{\prime \prime} \right. \notag \\ & \left. + \int _{z} ^{z ^{\prime}} \sin [pz] \sin [p (D - z ^{\prime \prime})] \sin [p z ^{\prime \prime}] \sin [p (D - z ^{\prime})] dz ^{\prime \prime} \right. \notag \\ & \left. + \int _{z ^{\prime}} ^{D} \sin [pz] \sin [p z ^{\prime}] \sin ^{2} [p (D - z ^{\prime \prime})] dz ^{\prime \prime} \right] , \label{E_parallel}
\end{align}
with the final result
\begin{align}
E _{\parallel} (z,z ^{\prime}) = & \frac{1}{4 p ^{3} \sin ^{2} [pD]} \bigg\{ \sin [p z _{>}] \sin [p z _{<}] \Big[ 2 p (D - z _{>} ) - \sin [ 2 p (D - z _{>} ) ] \Big] \label{E-Plates-FIN} \\[4pt] & + \sin[p (D - z _{>})] \sin[p z _{<}] \Big[ \sin[p (D - 2 z _{<})] - \sin[p (D - 2 z _{>})] -2 p (z _{>} - z _{<}) \cos [pD] \Big] \notag \\[4pt] & + \sin[p (D - z _{>}) ] \sin [p (D - z _{<})] \Big[ 2 p z _{<} - \sin [2 p z _{<}] \Big] \bigg\} . \notag
\end{align}
With the help of this result and the reduced GF (\ref{RedGFPlates}) we can compute the $M _{\parallel}$-function. We obtain
\begin{align}
M _{\parallel} (z,z ^{\prime}) = & \frac{\csc ^{2} [pD]}{32 p ^{5}} \bigg\{ 2 p z _{<} \sin[p z _{>}] \cos[p z _{<}] \Big[ 4 p D  - 2 p z _{>} + 2 p z _{>} \cos [2 p D ] + 3 \sin[ 2 p D ] \Big] \label{M-Plates-FIN} \\[4pt]   & - \csc [ p D ] \sin [ p z _{>}] \sin [ p z _{<} ] \Big[ \Big( 3 + 8 D ^{2} p ^{2} - p ^{2} ( z _{>} ^{2} + z _{<} ^{2}) \Big) \cos [ p D ] + \Big(- 3 + p ^{2} (z _{>} ^{2} + z _{<} ^{2}) \Big) \cos [ 3 D p ] \Big] \notag \\[4pt] & - \csc [ p D ] \sin [ p z _{>}] \sin [ p z _{<}] \Big( 6 p \big( 2 D - z _{>} + z _{>} \cos [ 2 p D ] \big) \sin [ p D ] \Big) \notag \\[4pt] & - 4 p z _{<} \cos [ p z _{>} ]  \cos [p z _{<} ] \sin [ p D ] \Big( 2 p z _{>} \cos [ p D ] + 3 \sin [ p D ] \Big) \notag \\[4pt] & - 2 \cos [p z _{>} ] \sin [p z _{<} ] \Big[ 3 + 4 D p ^{2} z _{>} - p ^{2}  z _{>} ^{2} - p ^{2} z _{<} ^{2} + \Big( - 3 + p ^{2} ( z _{>} ^{2} + z _{<} ^{2} ) \Big) \cos [ 2 p D ] + 3 p z _{>} \sin [ 2 p D] \Big] \bigg\} . \notag
\end{align}
On can further check that this function satisfies the boundary conditions $M _{\parallel} (0,z ^{\prime})  = M _{\parallel} (D,z ^{\prime})  = 0$, which are inherited from the standard boundary conditions on the conducting plates.

\subsection*{Infinite conducting plate}

Finally we consider the case of a single conducting planar plate located at $z = D > 0$. The reduced GF describing the propagation at the right hand side ($z,z^{\prime} > D$) is
\begin{align}
\mathfrak{g} _{\mid} (z,z ^{\prime}) = \frac{1}{p} \sin [p (z _{<} -D)] e ^{i p (z _{>} - D)} , \label{GFSinglePlate}
\end{align}
which is defined in the domain $\mathcal{D} _{\mid} = [D , \infty )$ and satisfies the boundary conditions $\mathfrak{g} _{\mid} (D,z ^{\prime}) = 0$ and $\mathfrak{g} _{\mid} (z \rightarrow \infty , z ^{\prime}) \sim e ^{ikz} $. Using Eq. (\ref{E-func-App}) we can further compute the associated $E _{\mid}$-function, i.e.
\begin{align}
E _{\mid} ( z , z ^{\prime} ) = \frac{1}{p ^{2}} \bigg[ & \int _{D} ^{z} \sin [p (z ^{\prime \prime} - D)] e ^{i p (z-D)} \sin [p (z ^{\prime \prime} - D)]  e ^{i p (z ^{\prime} - D)} dz ^{\prime \prime} \\ \nonumber + & \int _{z} ^{z ^{\prime}}  \sin [p (z - D)]  e^{i p (z ^{\prime \prime} - D)}   \sin [p (z ^{\prime \prime} - D)] e ^{i p (z ^{\prime}-D)} dz ^{\prime \prime} \\ \nonumber + & \int _{z ^{\prime}} ^{\infty} \sin[p (z - D)] e ^{ i p (z ^{\prime \prime} - D)} \sin[p (z ^{\prime} - D)] e ^{i p (z ^{\prime \prime} - D)}  d z ^{\prime \prime} \bigg],
\end{align}
and the final result is
\begin{align}
E _{\mid} (z,z ^{\prime}) = \frac{1}{4 p ^{3}} \Bigg\{ e ^{-i p (3 D- z_>)} \sin [p (D - z _{<})] \Big[ e ^{2 i pD} \big[ 1- 2 i p (z _{>} - z _{<} ) \big] -e^{2 i p z _{<}} \Big] \label{E-1plate} \\ + e ^{- i p (2 D - z - z _{<} )} \Big[ 2 p (z _{<} - D ) + \sin [2 p (D - z _{<})] \Big] \Bigg\} \notag .
\end{align}
We close this section with the $M _{\mid}$-function, which can be computed with the help of Eqs. (\ref{M-func-App}), (\ref{GFSinglePlate}) and (\ref{E-1plate}). The result is
\begin{align}
M _{\mid} ( z , z ^{\prime} ) &= \frac{1}{32 p ^{5}} \Bigg\{ e ^{- i p (4 D - z _{>} + z _{<} )} \Bigg[ e ^{2 i p ( D + z _{>})} \Big[ 2 p ( z _{>} - z _{<} ) + 3 i \Big] + 2 p e ^{4 i p D } ( z _{>} - z _{<} ) \Big[ 2 - i p ( z _{>} - z _{<}) \Big]  \Bigg] \label{M-1Plate} \\ \nonumber
& + e ^{ - i p (4 D - z _{>} + z _{<})} \Bigg[ e ^{4 i p z _{<}} (- 4 p D + 4 p z _{<} + 3 i ) - e ^{2 i p ( z _{>} + z _{<})} \Big[ 2 p (- 2 D + z _{>} + z _{<}) + 3 i \Big]  \Bigg] \\ \nonumber
& + e ^{ - i p (4 D - z _{>} + z _{<} )} \Bigg[ i e ^{2 i p ( D + z _{<} )} \Big[ - 3 + 2 p ( z _{>} - z _{<} ) \big[ p ( - 4 D + z _{>} + 3 z _{<} ) + 2 i \big] \Big]  \Bigg] \\ \nonumber
& - 2 e ^{ - i p (3 D - 2 z _{>} - z _{<} )} \sin [p (z _{>} - D)]  \Bigg[ e ^{2 i p (D - z _{<})} \Big[ - 3 + 2 i p ( z _{>} - z _{<}) \Big] + 4 i p D - 2 i p (z _{>} + z _{<} ) + 3 \Bigg] \\ \nonumber
& + 4 p e ^{- i p (2 D - z _{>} - z _{<} )} ( D - z _{<} ) \Big[ 2 i p ( D - z _{<} ) + \cos [2 p ( D - z _{<} )] + 2 \Big] \\
& + 2 e ^{- i p (2 D - z _{>} - z _{<} )} \sin[2 p ( D - z _{<} )] \Big[ - 3 - 2 i p ( D - z _{<} ) \Big]  \Bigg\} . \nonumber
\end{align}
The boundary conditions $E _{\mid} ( D , z ^{\prime} ) = M _{\mid} ( D , z ^{\prime} ) = 0$ can be directly verified.

\acknowledgments

We thank A. Kosteleck\'{y} and L. F. Urrutia for many valuable discussions, comments, and suggestions. This work has been partially supported by the project DGAPA-UNAM, Project No. IN-104815 and the CONACyT Project No. 237503. C. A. E. is supported by CONACyT Posdoctoral Grant No. 234745.


\begin{thebibliography}{99}

\bibitem{Noconmu} I. Mocioiu, M. Pospelov, and R. Roiban, Phys. Lett. B \textbf{489}, 390 (2000); S. M. Carroll \textit{et al.}, Phys. Rev. Lett. \textbf{87}, 141601 (2001); C. E. Carlson, C. D. Carone, and R. F. Lebed, Phys. Lett. B \textbf{518}, 201 (2001); A. Anisimov, T. Banks, M. Dine, and M. Graesser, Phys. Rev. D \textbf{65}, 085032 (2002).

\bibitem{Bworld} C. P. Burgess \textit{et al.}, JHEP \textbf{0203}, 043 (2002); A. R. Frey, JHEP \textbf{0304}, 012 (2003); J. M. Cline and L. Valc\'{a}rcel, JHEP \textbf{0403}, 032 (2004).

\bibitem{emergentG} C. Barcelo, S. Liberati, and M. Visser, Living Rev. Rel. \textbf{8}, 12 (2005).

\bibitem{modelos1} J. R. Ellis, N. E. Mavromatos, D. V. Nanopoulos, A. S. Sakharov and E. K. Sarkisyan, Astropart. Phys. \textbf{25}, 402 (2006); G. Amelino-Camelia, Phys. Lett. B \textbf{510}, 255 (2001); A. Connes and D. Kreimer, Commun. Math. Phys. \textbf{199}, 203 (1998).

\bibitem{modelos2} M. R. Douglas and N. A. Nekrasov,  Rev. Mod. Phys. \textbf{73}, 977-1029 (2001); L. F. Urrutia, Lect. Notes Phys. \textbf{702}, 299-345 (2006); P. Horava, Phys. Rev. D \textbf{79}, 084008 (2009); G. Amelino-Camelia, Phys. Lett. B \textbf{510}, 255-263 (2001); G. Amelino-Camelia, L. Freidel, J. Kowalski-Glikman and L. Smolin, Phys. Rev. D \textbf{84}, 084010 (2011).

\bibitem{modelos3}  C. A. Escobar and L. F. Urrutia, Phys. Rev. D \textbf{92}, 025042 (2015); C. A. Escobar and L. F. Urrutia, Phys. Rev. D \textbf{92}, 025013 (2015); C. A. Escobar and L. F. Urrutia, Eur. Phys. Lett. \textbf{106} 31002 (2014); C. A. Escobar and M. A. G. Garcia, Phys. Rev. D \textbf{92}, 025034 (2015); Y. Bonder and C. A. Escobar, Phys. Rev. D \textbf{93}, 025020 (2016).

\bibitem{quantum1}  R. Gambini and J. Pullin, Phys. Rev. D \textbf{59}, 124021 (1999); J. Alfaro, H. A. Morales-T\'{e}cotl, and L. F. Urrutia, Phys. Rev. D \textbf{65}, 103509 (2002); J. Alfaro, H. A. Morales-T\'{e}cotl, and L. F. Urrutia, Phys. Rev. Lett. \textbf{84}, 2318 (2000); S. M. Carroll, J. A. Harvey, V. A. Kosteleck\'{y}, C. D. Lane, and T. Okamoto, Phys. Rev. Lett. \textbf{87}, 141601 (2001).

\bibitem{quantum2} G. Amelino-Camelia, J. R. Ellis, N. E. Mavromatos, D. V. Nanopoulos, and S. Sarkar, Nature \textbf{393}, 763 (1998); J. Magueijo and L. Smolin, Phys. Rev. D \textbf{67} 044017 (2003).


\bibitem{String2} V. A. Kosteleck\'{y} and S. Samuel, Phys. Rev. D \textbf{39}, 683 (1989); V. A. Kosteleck\'{y} and R. Potting, Nucl. Phys. B \textbf{359}, 545 (1991); K. Hashimoto and M. Murata, PTEP \textbf{2013}, 043B01 (2013).

\bibitem{QG1} F. R. Klinkhamer and C. Rupp, Phys. Rev. D \textbf{70}, 045020 (2004); A. Mart\'{i}n-Ruiz, Phys. Rev. D \textbf{90}, 125027 (2014); A. Mart\'{i}n-Ruiz \textit{et al.}, Rev. Mex. F\'{i}s. \textbf{61}, 182 (2015); A. Mart\'{i}n-Ruiz, A. Frank, and L. F. Urrutia, Phys. Rev. D \textbf{92}, 045018 (2015).


\bibitem{SME} D. Colladay and V. A. Kosteleck\'{y}, Phys. Rev. D \textbf{55}, 6760 (1997); Phys. Rev. D \textbf{58}, 116002 (1998).

\bibitem{SME2} V. A. Kosteleck\'{y}, Phys. Rev. D \textbf{69}, 105009 (2004).

\bibitem{Strings} V. A. Kosteleck\'{y} and S. Samuel, Phys. Rev. D \textbf{40}, 1886 (1989); Phys. Rev. D \textbf{39}, 683 (1989); Phys. Rev. Lett. \textbf{63}, 224 (1989).

\bibitem{Ren1} V.A. Kosteleck\'{y}, C. Lane, and A. Pickering, Phys. Rev. D \textbf{65}, 056006 (2002).
\bibitem{Ren2} V.A. Kosteleck\'{y} and A. Pickering, Phys. Rev. Lett. \textbf{91}, 031801 (2003).
\bibitem{Ren3}  D. Colladay and P. McDonald, Phys. Rev. D \textbf{75}, 105002 (2007); Phys. Rev. D \textbf{77}, 085006 (2008); Phys. Rev. D \textbf{79}, 125019 (2009).

\bibitem{Causality} V.A. Kosteleck\'{y} and R. Lehnert, Phys. Rev. D \textbf{63}, 065008 (2001).

\bibitem{Light1} V.A. Kosteleck\'{y} and M. Mewes, Phys. Rev. D \textbf{66}, 056005 (2002).

\bibitem{Light2} Don Colladay and V.A. Kosteleck\'{y}, Phys. Lett. B \textbf{511}, 209-217 (2001); M. Perez, JHEP \textbf{0104} 32 (2001); V.A. Kosteleck\'y, C.D. Lane, and A.G. Pickering, Phys. Rev. D \textbf{65}, 056006 (2002); M. Cambiaso, R. Lehnert, and R. Potting, Phys. Rev. D \textbf{90}, 065003 (2014); M. Cambiaso, R. Lehnert, and R. Potting, Phys. Rev. D \textbf{85}, 085023 (2012); B. Altschul, Phys. Rev. D \textbf{75}, 105003 (2007); M. Schreck, Phys. Rev. D \textbf{86}, 065038 (2012); M. Schreck, Phys. Rev. D \textbf{89}, 085013 (2014); F.R. Klinkhamer and M. Schreck, Nucl. Phys. B \textbf{848} 90-107 (2011); R. Casana, M.M. Ferreira Jr., A.R. Gomes, and P.R. Pinheiro, Eur. Phys. J. C \textbf{62}, 573 (2009); F.R. Klinkhamer and M. Risse, Phys. Rev. D \textbf{77}, 016002 (2008); F.R. Klinkhamer and M. Risse, Phys. Rev. D \textbf{77}, 117901(A) (2008); A. Kobakhidze and B.H.J. McKellar, Phys. Rev. D \textbf{76}, 093004 (2007).
\bibitem{Light3} P. Wolf, S. Bize, A. Clairon, G. Santarelli, M.E. Tobar, and A.N. Luiten, Phys. Rev. D \textbf{70}, 051902(R) (2004); P.L Stanwix, M.E. Tobar, P. Wolf, C.R. Locke, and E.N. Ivanov, Phys. Rev. D \textbf{74}, 081101(R) (2006); V.A. Kosteleck\'{y} and M. Mewes, Phys. Rev. Lett. \textbf{87}, 251304 (2001).

\bibitem{Cherenkov1} D. Colladay, P. McDonald and R. Potting, Phys. Rev. D \textbf{93}, 125007 (2016).

\bibitem{Cherenkov2}  V.A. Kosteleck\'{y} and J. D. Tasson, Phys. Lett. B 749, 551 (2015).

\bibitem{Cherenkov3} R. Lehnert and R. Potting, Phys. Rev. Lett. \textbf{93}, 110402 (2004); Phys. Rev. D \textbf{70}, 125010(2004) [Phys. Rev. D \textbf{70}, 129906 (2004)]; C. Kaufhold and F. R. Klinkhamer, Phys. Rev. D \textbf{76}, 025024 (2007).

\bibitem{CFJ} S. M. Carroll, G. B. Field, and R. Jackiw, Phys. Rev. D \textbf{41}, 1231 (1990).

\bibitem{Casimir} H. B. G. Casimir, Proc. K. Ned. Akad. Wet. 51, 793 (1948).

\bibitem{Bressi} G. Bressi \textit{et al}, Phys. Rev. Lett. \textbf{88}, 041804 (2002).

\bibitem{Quach} J. Q. Quach, Phys. Rev. Lett. \textbf{114}, 081104 (2015).

\bibitem{AdS} R. C. Myers, Phys. Rev. D \textbf{60}, 046002 (1999).

\bibitem{CE-TI2} A. G. Grushin and A. Cortijo, Phys. Rev. Lett. \textbf{106}, 020403 (2011).

\bibitem{CE-TI}A. G. Grushin, P. Rodriguez-Lopez and A. Cortijo, Phys. Rev. B \textbf{84}, 045119 (2011); P. Rodriguez-Lopez and A. G. Grushin, Phys. Rev. Lett. \textbf{112}, 056804 (2014); A. Mart\'{i}n-Ruiz, M. Cambiaso and L. F. Urrutia, Eur. Phys. Lett. \textbf{113}, 60005 (2016).

\bibitem{CE-MM} R. Zhao, J. Zhou, Th. Koschny, E. N. Economou, and C. M. Soukoulis, Phys. Rev. Lett. \textbf{103}, 103602 (2009); F. S. S. Rosa, D. A. R. Dalvit, and P. W. Milonni, Phys. Rev. A \textbf{78}, 032117 (2008).

\bibitem{Martin-Escobar} A. Mart\'{i}n-Ruiz and C. A. Escobar, Phys. Rev. D \textbf{94}, 076010 (2016).

\bibitem{Mariana} M. Frank and I. Turan, Phys. Rev. D \textbf{74}, 033016 (2006).

\bibitem{Kharlanov} O. G. Kharlanov and V. Ch. Zhukovsky, Phys. Rev. D \textbf{81}, 025015 (2010).

\bibitem{cotas} V.A. Kosteleck\'y and N. Russell, Rev. Mod. Phys. \textbf{83}, 11-31 (2011).

\bibitem{TI} A. Mart\'{i}n-Ruiz, M. Cambiaso and L. F. Urrutia, Phys. Rev. D \textbf{92}, 125015 (2015); A. Mart\'{i}n-Ruiz, M. Cambiaso and L. F. Urrutia, Phys. Rev. D \textbf{93}, 045022 (2016); A. Mart\'{i}n-Ruiz, M. Cambiaso and L. F. Urrutia, Phys. Rev. D \textbf{94}, 085019 (2016).

\bibitem{WSM} X. L. Qi and S. C. Zhang, Rev. Mod. Phys. \textbf{83}, 1057 (2011), M. Z. Hasan and C. L. Kane, Rev. Mod. Phys. \textbf{82}, 3045 (2010).

\bibitem{brasilenos} R. Casana, M. M. Ferreira Jr. and C. E. H. Santos, Phys. Rev. D \textbf{78}, 025030 (2008).

\bibitem{Milton} K. A. Milton, \textit{The Casimir effect: Physical Manifestation of Zero-Point Energy} (World Scientific, Singapore, 2001); M. Bordag, G. L. Klimchitskaya, U. Mohideen and V. M. Mostepanenko, \textit{Advances in Casimir effect} (Oxford University Press, Great Britain, 2009).

\bibitem{Candelas} D. Deutsch and P. Candelas, Phys. Rev. D \textbf{20}, 3063 (1979).

\bibitem{Sopova} V. Sopova and L. H. Ford, Phys. Rev. D \textbf{66}, 045026 (2002); Phys. Rev. D \textbf{72}, 033001 (2005).


\end{thebibliography}
\end{document}